\documentclass[preprint2]{aastex63} % arxiv version

\graphicspath{{./}{figures/}}

\newcommand{\Teff}{\mbox{$T_{\rm eff}$}}

\newcommand{\kms}{\mbox{km\,s$^{-1}$}}
\newcommand{\vsini}{\mbox{$v\sin i$}}

\newcommand{\dsct}{\mbox{$\delta$~Scuti}}

\newcommand{\gdor}{\mbox{$\gamma$~Doradus}}

\newcommand{\cd}{\mbox{d$^{-1}$}}

\newcommand{\Dnu}{\mbox{$\Delta\nu$}}

\newcommand{\numax}{\mbox{$\nu_{\rm max}$}}
\renewcommand{\numax}{\texorpdfstring{\mbox{$\nu_{\rm max}$}}{numax}}
\newcommand{\bprp}{\mbox{$G_{BP}-G_{RP}$}}
\newcommand{\alphaMLT}{\mbox{$\alpha_{\rm MLT}$}}
\newcommand{\kepler}{{\em Kepler\/}}
\renewcommand{\kepler}{Kepler}
\newcommand{\tess}{{\em TESS\/}}
\renewcommand{\tess}{TESS}

\newcommand{\new}[1]{{\color{red}{#1}}}
\renewcommand{\new}[1]{\relax{#1}}

\newcommand{\newtwo}[1]{{\color{red}{#1}}}
\renewcommand{\newtwo}[1]{\relax{#1}}

\newcommand{\newthree}[1]{{\color{red}{#1}}}
\renewcommand{\newthree}[1]{\relax{#1}}

\usepackage{longtable}
\usepackage{booktabs} 
\graphicspath{{./}{figures/}}
\newif\ifarxiv
\arxivtrue % for arxiv or journal version
\usepackage{CJKutf8}
\newcommand{\CNnames}[1]{{\begin{CJK}{UTF8}{gbsn}~(#1)~\end{CJK}}}

\begin{document}

\title{\textit{TESS} observations of the Pleiades cluster: a nursery for $\delta$~Scuti stars}
% TESS observations of the Pleiades cluster: a nursery for delta Scuti stars

%% Note that the corresponding author command and emails has to come
%% before everything else. Also place all the emails in the \email
%% command instead of using multiple \email calls.
\correspondingauthor{Timothy R. Bedding}
\email{tim.bedding@sydney.edu.au}

%Timothy R. Bedding, Simon J. Murphy, Courtney Crawford, Daniel R. Hey, Daniel Huber, Hans Kjeldsen, Yaguang Li, Andrew W. Mann, Guillermo Torres, Timothy R. White and George Zhou

\author[0000-0001-5222-4661]{Timothy R. Bedding}
\affiliation{Sydney Institute for Astronomy, School of Physics, University of Sydney NSW 2006, Australia}

\author[0000-0002-5648-3107]{Simon J. Murphy}
\affiliation{Centre for Astrophysics, University of Southern Queensland, Toowoomba, QLD 4350, Australia}

\author[0000-0002-7654-7438]{Courtney Crawford}
\affiliation{Sydney Institute for Astronomy, School of Physics, University of Sydney NSW 2006, Australia}

\author[0000-0003-3244-5357]{Daniel R. Hey}
\affiliation{Institute for Astronomy, University of Hawai`i, Honolulu, HI 96822, USA}

\author[0000-0001-8832-4488]{Daniel Huber}
\affiliation{Institute for Astronomy, University of Hawai`i, Honolulu, HI 96822, USA}

\author[0000-0002-9037-0018]{Hans Kjeldsen}
\affiliation{Stellar Astrophysics Centre, Department of Physics and Astronomy, Aarhus University, DK-8000 Aarhus C, Denmark}

\author[0000-0003-3020-4437]{Yaguang Li\CNnames{李亚光}}
\affiliation{Sydney Institute for Astronomy, School of Physics, University of Sydney NSW 2006, Australia}

\author[0000-0003-3654-1602]{Andrew W. Mann}
\affiliation{Department of Physics and Astronomy, University of North Carolina at Chapel Hill, Chapel Hill, NC, USA}

\author[0000-0002-5286-0251]{Guillermo Torres}
\affiliation{Center for Astrophysics $\vert$ Harvard \& Smithsonian, 60 Garden Street, Cambridge, MA 02138, USA}

\author[0000-0002-6980-3392]{Timothy R. White}
\affiliation{Sydney Institute for Astronomy, School of Physics, University of Sydney NSW 2006, Australia}

\author[0000-0002-4891-3517]{George Zhou}
\affiliation{Centre for Astrophysics, University of Southern Queensland, Toowoomba, QLD 4350, Australia}

\begin{abstract}
We studied 89 A- and F-type members of the Pleiades open cluster, including five escaped members.  We measured projected rotational velocities (\vsini) for 49 stars and confirmed that stellar rotation causes a broadening of the main sequence in the color-magnitude diagram. Using time-series photometry from NASA's \tess\ Mission (plus one star observed by \kepler/K2), we detected \dsct\ pulsations in 36 stars.  
The fraction of Pleiades stars in the middle of the instability strip that pulsate is unusually high (over 80\%), and their range of effective temperatures agrees well with theoretical models.  On the other hand, the characteristics of the pulsation spectra are varied and do not correlate with stellar temperature, calling into question the existence of a useful \numax\ relation for \dsct{}s, at least for young main-sequence stars.  By including \dsct\ stars observed in the \kepler\ field, we show that the instability strip is shifted to the red with increasing distance by interstellar reddening.  
Overall, this work demonstrates the power of combining observations with Gaia and \tess\ for studying pulsating stars in open clusters.
\end{abstract}

%We studied 89 A- and F-type members of the Pleiades open cluster, including five escaped members.  We measured projected rotational velocities (v sin i) for 49 stars and confirmed that stellar rotation causes a broadening of the main sequence in the color-magnitude diagram. Using time-series photometry from NASA's TESS Mission (plus one star observed by Kepler/K2), we detected delta Scuti pulsations in 36 stars.  The fraction of Pleiades stars in the middle of the instability strip that pulsate is unusually high (over 80%), and their range of effective temperatures agrees well with theoretical models.  On the other hand, the characteristics of the pulsation spectra are varied and do not correlate with stellar temperature, calling into question the existence of a useful nu_max relation for delta Scutis, at least for young main-sequence stars.  By including delta Scuti stars observed in the Kepler field, we show that the instability strip is shifted to the red with increasing distance by interstellar reddening.  Overall, this work demonstrates the power of combining observations with Gaia and TESS for studying pulsating stars in open clusters.


%%%%%%%%%%%%%%%%%%%%%%%%%%%%%%%%%%%%%%%%%%%%%%%%%%%%%%%%%%%%%%%%
\section{Introduction} \label{sec:intro}

Explaining the \newtwo{details of} excitation and mode selection in \dsct\ stars is one of the major unsolved challenges in stellar pulsations  (see reviews by \citealt{Goupil++2005, Handler2009, Lenz2011, Guzik2021, Kurtz2022}).  Why do only a subset of stars in the instability strip show \dsct\ pulsations?  And how can two stars occupy essentially the same position in the H--R diagram but only one show pulsations?

One obvious explanation is that some stars have pulsations too weak to be detected.  However, CoRoT and \kepler\ pushed the detection threshold down to extremely low levels \citep{Balona++2015, Michel2017, Bowman+Kurtz2018, Guzik2021}, and it is still the case that only about half the stars in the central part of the instability strip are pulsating \citep{Murphy2019}.  

Another possible factor is chemical composition.  Stars with different metallicities can pass through a given location in the H--R diagram at different ages, and with different opacities in the driving zone. This will affect pulsations driven by the $\kappa$ (opacity) mechanism \citep{guzik++2018}, and even more so for chemically peculiar stars \citep{murphy++2015a,guzik++2021}, so
it is likely that chemical composition is part of the explanation.  But even in open clusters, which are assumed to have a uniform metallicity \citep{DeSilva++2006, Sestito++2007, Bovy2016}, the pulsator fraction is much less than 1.  In the Pleiades, for example, \citet{Breger1972-pleiades} found four \dsct\ stars and three decades later, that number still only stood at six \citep{Koen++1999, Li++2002, Fox-Machado++2006}.  \kepler/K2 observed five of these in short-cadence (1 minute) mode \citep{Murphy++2022-K2-Pleiades}, and the long-cadence (30 minutes) data hinted at more variables \citep{Rebull++2016-II}.  

% A third factor is rotation.  Many A/F stars rotate rapidly, which affects their evolution, their observed properties (through gravity darkening) and their pulsations.  

NASA's \tess\ Mission \citep{Ricker++2015} is producing high-precision, rapid-cadence light curves over most of the sky, opening up new possibilities for studying large samples of \dsct\ stars \citep[e.g.,][]{Antoci++2019, Balona+Ozuyar2020, Barcelo-Forteza++2020, Bedding++2020, Murphy++2020-lambda-boo}. In this {\em Letter}, we use data from Gaia and \tess\ to perform the most detailed search to date for \dsct\ pulsators in the Pleiades open cluster (Messier~45).

% \TB{Discuss \dsct\ stars in other open clusters (Praesepe, Hyades, M67 blue stragglers)? \citep{Breger1972-open-clusters}}

%%%%%%%%%%%%%%%%%%%%%%%%%%%%%%%%%%%%%%%%%%%%%%%%%%%%%%%%%%%%%%%%
\section{Sample selection and Gaia photometry} \label{sec:sample}

% Mention previous papers on membership: \citet{Olivares++2018}

We selected an initial list of likely Pleiades members using Gaia DR2 astrometry and the BANYAN-$\Sigma$ code \citep{BanyanSigma}. We used the default Pleiades parameters and did not include any radial velocity information (to avoid biasing against binaries). We selected all stars with BANYAN membership probabilities above 90\% and Gaia colors $0.0 < G_{BP}-G_{RP} < 0.7$, which correspond approximately to spectral types in the range A0\,V to F8\,V.  This gave a list of 83 stars.  
We note that our membership selection was not altered by updating to Gaia DR3. 

We also included five stars listed by \citet{Heyl++2022} as escaped Pleiades members (HD~17962, HD~20655, HD~21062, HD~23323 and HD~34027). These stars are too distant from the Pleiades core to have been included in our BANYAN-$\Sigma$ selection. We note that a cross-check of the G and early K dwarfs in the \citet{Heyl++2022} sample with {\it TESS} indicated most of the suggested escapees have $<10$\,day rotation periods, which is consistent with expectations for Pleiades membership \citep{Curtis_stall}. 

Our final sample of 89 stars is listed in Table~\ref{tab:sample}.  
V1229~Tau (HD~23642) is a well-studied eclipsing and spectroscopic binary that consists of two A-type stars with an orbital period of 2.4611\,days \citep[see][and references therein]{Groenewegen++2007}.  
Both components are A-type stars, so we have listed them separately in the table (see Sec.~\ref{sec:v1229-tau} for details).

Ten stars in Table~\ref{tab:sample} are named variables (column~1).  These include the six \dsct\ stars previously known from ground-based observations (V534 Tau, V624 Tau, V647 Tau, V650 Tau, V1187 Tau and V1228 Tau; 
\citealt{Breger1972-pleiades, Koen++1999, Li++2002}), together with two \gdor\ stars (V1210 Tau and V1225 Tau; \citealt{Martin+Rodriguez2000}) and both members of the eclipsing binary V1229~Tau (HD~23642).

\begin{figure*}
\includegraphics[width=0.5\linewidth]{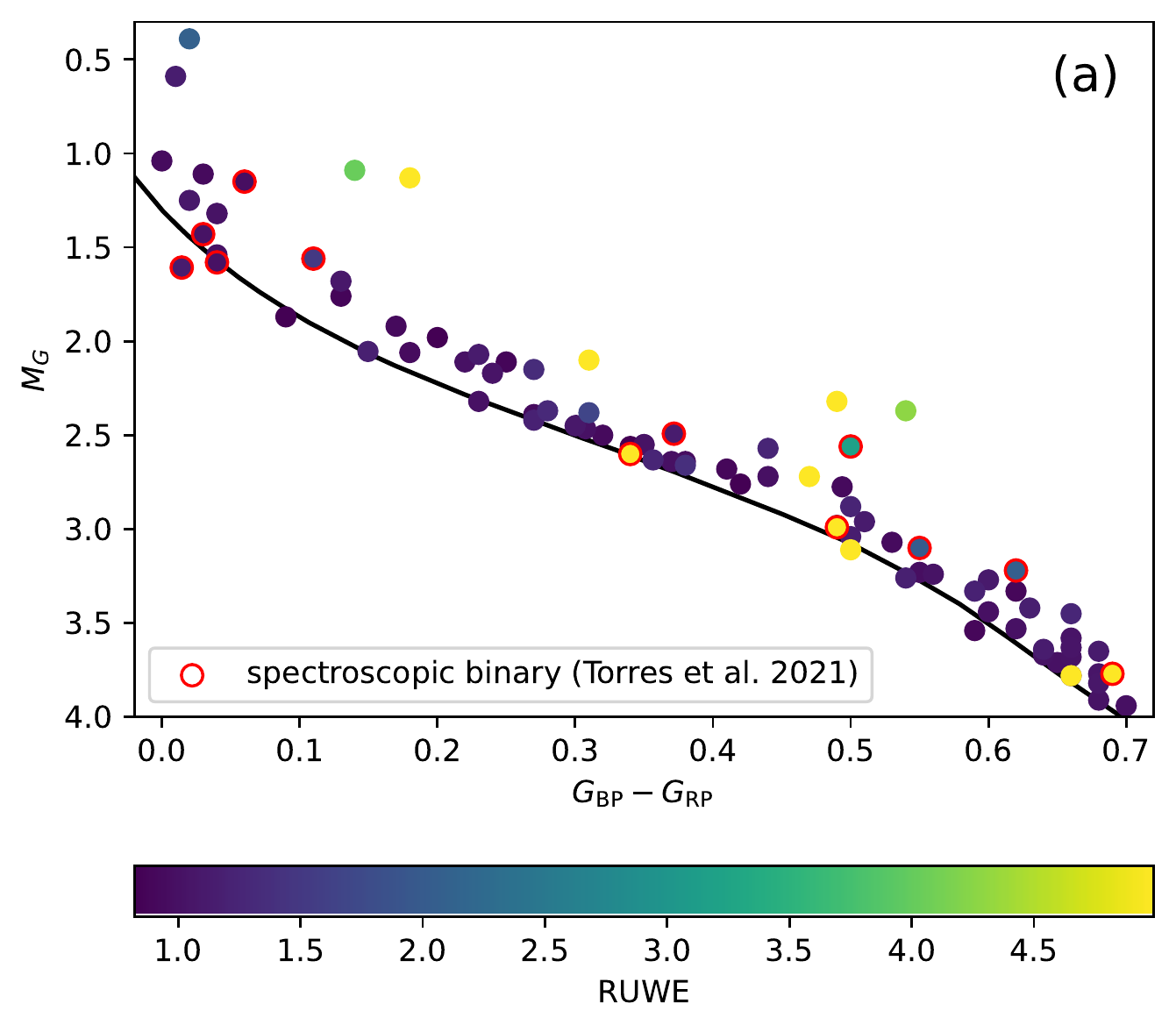}
\includegraphics[width=0.5\linewidth]{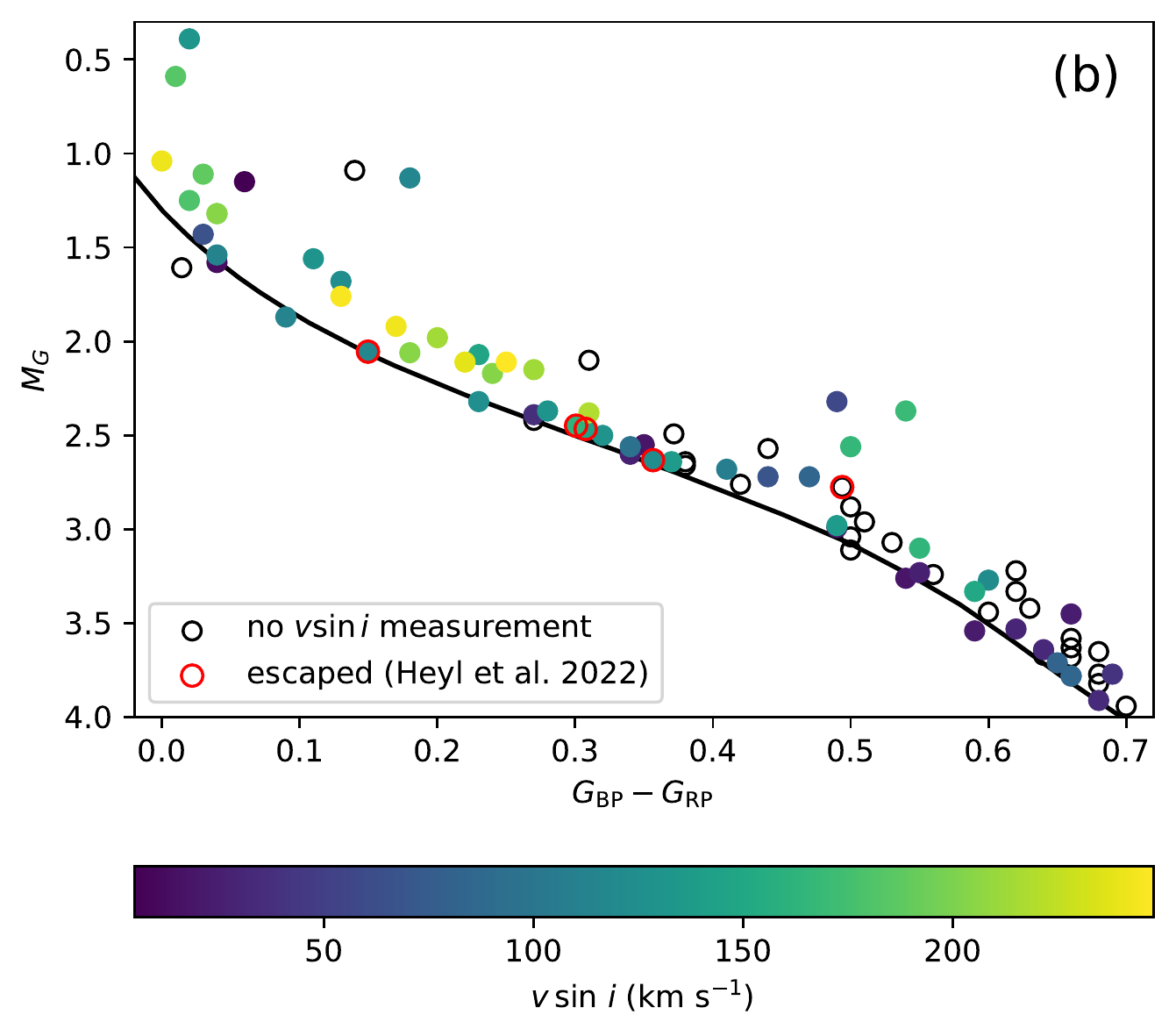}
\caption{CMD of 89 A and F stars in the Pleiades, based on photometry and parallaxes from Gaia DR3.  Photometry has not been corrected for extinction and reddening.  (a)~Stars are color coded by RUWE (clipped at $\mbox{RUWE}=5$, although some stars have greater values), and red circular outlines indicate spectroscopic binaries \citep{Torres++2021}; (b)~Stars are color-coded by \vsini\ (see Table~\ref{tab:sample}) and red circular outlines indicate five stars listed as escaped members by \citet{Heyl++2022}. The black line in both panels is a PARSEC isoschrone, corrected for extinction and reddening (see Section~\ref{sec:sample}).}
\label{fig:CMD}
\end{figure*}

The photometry in Table~\ref{tab:sample} (columns~5--7) is based on magnitudes and parallaxes from Gaia DR3 \citep{Gaia-Brown++2021,Lindegren++2021,Riello++2021}.  
In Fig.~\ref{fig:CMD} we show the color-magnitude diagram (CMD) of the sample.  No correction for extinction or reddening was made.
We have included a PARSEC isochrone \citep{Marigo++2017}, with a metallicity of $Z=0.017$ and an age of 110\,Myr, which are appropriate for the Pleiades \citep{Gaia-Babusiaux++2018}. We shifted the isochrone to account for  extinction and reddening, using values of $A_G=0.11$ and $E(\bprp) = 0.055$ \citep{Andrae++2018}.

Some of the spread in the cluster main sequence is from binarity.  The twelve red circular outlines in Fig~\ref{fig:CMD}a mark known spectroscopic binaries from the list compiled by \citet{Torres++2021}, and some of these clearly lie above the cluster sequence.  Figure\ref{fig:CMD}a also shows several stars above the main sequence with values of Gaia RUWE (renormalised unit weight error) significantly greater than 1.0, indicating they are likely to be binaries \citep{Evans2018,Belokurov++2020}. 
\new{Note that stars with high RUWE values can still provide useful parallaxes, although generally with larger uncertainties \citep[e.g.,][]{El-Badry-2021, Lindegren++2021, Maiz-Apellaniz++2021}.  As a check, we examined the \texttt{fidelity\_v2} diagnostic calculated by \citet{Rybizki++2022} and found it to have a value of 1.0 for all stars in our sample, indicating the astrometric solutions are reliable.
}

We conclude that stars well above the cluster sequence have high RUWE, are spectroscopic binaries, or both. 
The remaining spread in the observed sequence can be attributed to the presence of rapid rotators, as discussed in the next section.

% In Torres et al 2021:
% HD 23792 is L (long-term trend)
% HD 23247 is L, but too red to be in our sample (BPRP=0.65)
% only one star is AST (astrometric), and that is too faint to be in our sample

%%%%%%%%%%%%%%%%%%%%%%%%%%%%%%%%%%%%%%%%%%%%%%%%%%%%%%%%%%%%%%%%
\section{Projected rotational velocities}
\label{sec:vsini}

We have measured \vsini\ for 49 stars in our sample using spectra collected by the Center for Astrophysics (CfA) survey \citep{Torres2020, Torres++2021}.  These were gathered with the Tillinghast Reflector \'Echelle Spectrograph (TRES), a high-resolution ($R=44,000$) fiber-fed \'echelle mounted on the 1.5m reflector at the Fred Lawrence Whipple Observatory, Arizona. Following \citet{Zhou++2018}, we extracted line profiles from each spectrum via a least-squares deconvolution \citep{Donati++1997} against a synthetic non-rotating ATLAS9 template \citep{Castelli+Kurucz2003}. The broadening profile was then modeled as a combination of kernels describing the effects of rotational, macroturbulent, and instrumental broadening \citep{Gray2005}.  The resulting \vsini\ values are listed in column~10 of Table~\ref{tab:sample} and are indicated as source~1 in column~11.
For an additional 10 stars, we used \vsini\ measurements from Gaia RVS spectra \citep{Creevey++2022}, which are indicated as source~2 in the table.  By way of validation, we note there is good consistency for 15 stars with \vsini\ measurements from both sources.
\new{We also note that the distribution of our \vsini\ measurements is similar to that of A-type stars in general \citep[e.g.,][]{Royer++2007, Zorec+Royer2012}, so that we can consider the Pleiades to be representative of the broader population.}

The color-magnitude diagram in Figure~\ref{fig:CMD}b is color-coded by \vsini.  It is well-known that rotation causes stars to move in the CMD \citep{PerezHernandez++1999, Fox-Machado++2006, EspinosaLara+Rieutord2011, Lipatov+Brandt2020, Wang++2022, Malofeeva++2023}.  
This is at least partly responsible for the extended main-sequence turn-offs seen in the CMDs of young and intermediate-age clusters \citep{Bastian+deMink2009, Yang++2013, Brandt+Huang2015, Goudfrooij++2017, 
%Bastian++2018, Marino++2018, 
Gossage++2019, Sun++2019, deJuanOvelar++2020, Kamann++2020, Chen-Jing++2022, He++2022}.
Rotation does not only affect the turn-off, but also broadens the main sequence itself, and we are seeing good evidence for this in the Pleiades in Fig.~\ref{fig:CMD}b.

%%%%%%%%%%%%%%%%%%%%%%%%%%%%%%%%%%%%%%%%%%%%%%%%%%%%%%%%%%%%%%%%
\section{\tess\ Observations and analysis}

\begin{figure*}
\begin{center}
\includegraphics[width=0.9\linewidth]{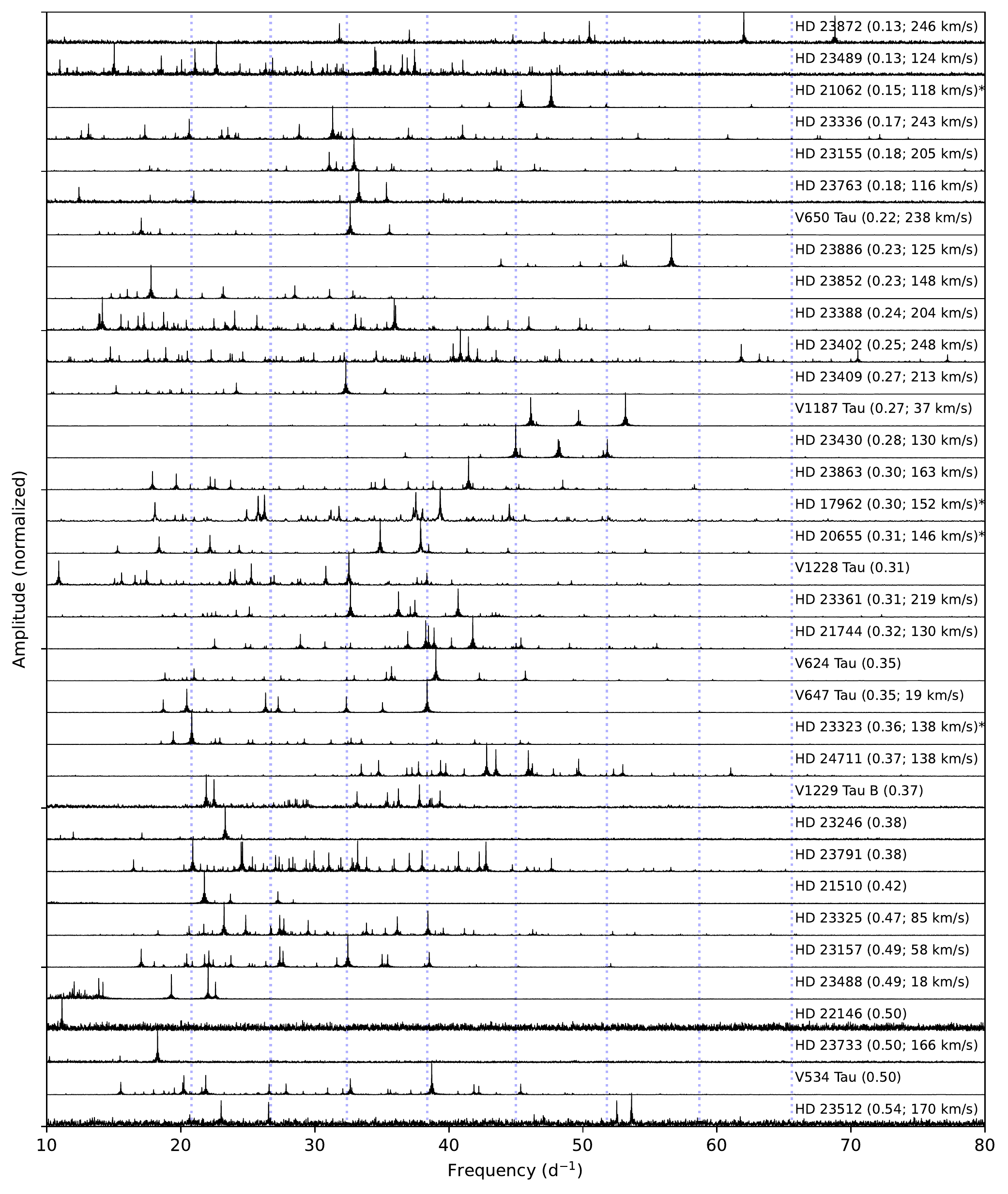}
\end{center}
\caption{Pulsation spectra of 35 \dsct\ stars in the Pleiades observed with \tess. Stars are ordered according to the Gaia \bprp\ color index, whose values are given in parentheses, together with \vsini\ (if available). Asterisks indicate four stars listed by \citet{Heyl++2022} as escaped members.
\newtwo{To help guide the eye, the blue dotted lines show approximate frequencies of the first 8 radial modes (see Sec.~\ref{sec:numax}). \newthree{These are based on the observed frequencies in the star V647~Tau, which has a particularly regular spectrum \citep[][Table~4]{Murphy++2022-K2-Pleiades}.}}
}
\label{fig:amp-spectra}
\end{figure*}

Observations with \tess\ are made in 27-d sectors \citep{Ricker++2015}.
%During the first three years of the mission the observations covered about 85\% of the sky, but this excluded the ecliptic plane.  
The Pleiades were observed in the fourth year of the mission, in Sectors 42--44 (2021 August 20 to November 6).  
Most Pleiades stars have \tess\ data in all three sectors and a few were also observed in Sectors 18 or 19.  All the stars in our sample except two have \tess\ observations with 120-s cadence, and we used the {\tt lightkurve} package \citep{lightkurve2018} to download the PDCSAP\footnote{Pre-search Data Conditioning Simple Aperture Photometry} light curves that were provided by the SPOC (Science Processing Operations Center).  The first exception was HD~23479.  For this star we extracted a light curve from the \tess\ full-frame images (10-min cadence), which showed no evidence for \dsct\ pulsations.\footnote{The light curve for HD~23479 was contaminated by oscillations from HD~23463 (separation 39\,arcsec), which is a red giant whose parallax and proper motion show it to be currently passing through the Pleiades cluster.  }  
The second exception was HD~23028, which fell just off the edge of the detector and  is the only star in our sample with no \tess\ observations.  For this star, \kepler/K2 long-cadence (30-min) observations show pulsations and we have included it as a \dsct\ star in the table and figures (apart from Fig.~\ref{fig:amp-spectra}).

% We used the \tess\ light curves to carry out a search for \dsct\ pulsations for all stars in our sample.

In total, we detected \dsct\ pulsations in 36 of the stars in our sample, as flagged in Table~\ref{tab:sample} (column~12).  These include the six previously known from ground-based observations (V534 Tau, V624 Tau, V647 Tau, V650 Tau, V1187 Tau and V1228 Tau), plus 30 additional detections.  For all detections, the amplitude spectrum showed several clear peaks at least 10 times the mean noise level (and often much higher), while the non-detections showed no significant peaks above about 4 times the noise.
\new{The last two columns of Table~\ref{tab:sample} show the frequency and amplitude of the strongest mode in each \dsct\ star (measured in the range 10 to 80\,\cd).}

The amplitude spectra for the 35 \dsct\ stars observed by \tess\ are shown in Fig.~\ref{fig:amp-spectra}, ordered according to Gaia \bprp.  The detections include four of the five escaped members \citep{Heyl++2022}, and the similarity of those oscillation spectra to confirmed members of the Pleiades lends support to their status as escaped members.

\subsection{V1229~Tau (HD~23642)}
\label{sec:v1229-tau}

V1229~Tau is a well-studied eclipsing and spectroscopic binary, with a period of 2.4611\,d \citep[see][and references therein]{Groenewegen++2007}.  Both components are A-type stars, and so we have treated them separately.

The Gaia photometry ($G=6.82$ and $\bprp = 0.10$) measures the combined light of the system.  In order to plot both components separately, we have estimated values in the table using the published effective temperatures ($9750 \pm 250$\,K and $7600 \pm 400$\,K; \citealt{Southworth++2005}) and a luminosity ratio of $0.355 \pm 0.035$ \citep{David++2016}.  The photometry given in columns 5--7 of Table~\ref{tab:sample} are estimates if the two components were measured separately, also taking into account the reddening and extinction of the cluster.  In Fig.~\ref{fig:CMD-dsct}a, the black circular outlines show (from left to right) the A component, the combined system, and the B component.  

In addition to the eclipses, the \tess\ light curve shows high-frequency \dsct\ pulsations.  The amplitude spectrum in Fig.~\ref{fig:amp-spectra} was made after fitting and subtracting an eclipse model.  Given the colors of the components (see Fig.~\ref{fig:CMD-dsct}a), it is reasonable to conclude that the pulsations occur in the B component.  To verify this, we examined the scatter in the time series after fitting and removing the five highest peaks in the amplitude spectrum.  We found the scatter to be reduced everywhere in this prewhitened light curve, but the reduction was less during secondary eclipses because the five-peak fit is a poorer fit when part of the pulsating star is being eclipsed (note that the inclination of the system is about 78$^\circ$ and the eclipses are not total; \citealt{David++2016}). We can therefore confirm that it is the secondary component (V1229~Tau~B) that is undergoing pulsations. 

\citet{Chen-Xinghao++2022} noted pulsations in V1229~Tau (which they referred to as TIC~125754991) and suggested that it is hotter than typical \dsct\ stars, because they assumed the primary is the pulsator.  Once we accept that the secondary is the pulsating component, this star becomes typical. A more detailed study of the pulsations of V1229~Tau (HD~23642) using the \tess\ light curve has been made by \citet{Southworth++2023}.

%From the \echelle, the mostly likely large separation is $\Dnu = 6.20\,\cd$.

%%%%%%%%%%%%%%%%%%%%%%%%%%%%%%%%%%%%%%%%%%%%%%%%%%%%%%%%%%%%%%%%
\section{The \texorpdfstring{$\delta$}{delta} Scuti instability strip}

\begin{figure}
\includegraphics[width=\linewidth]{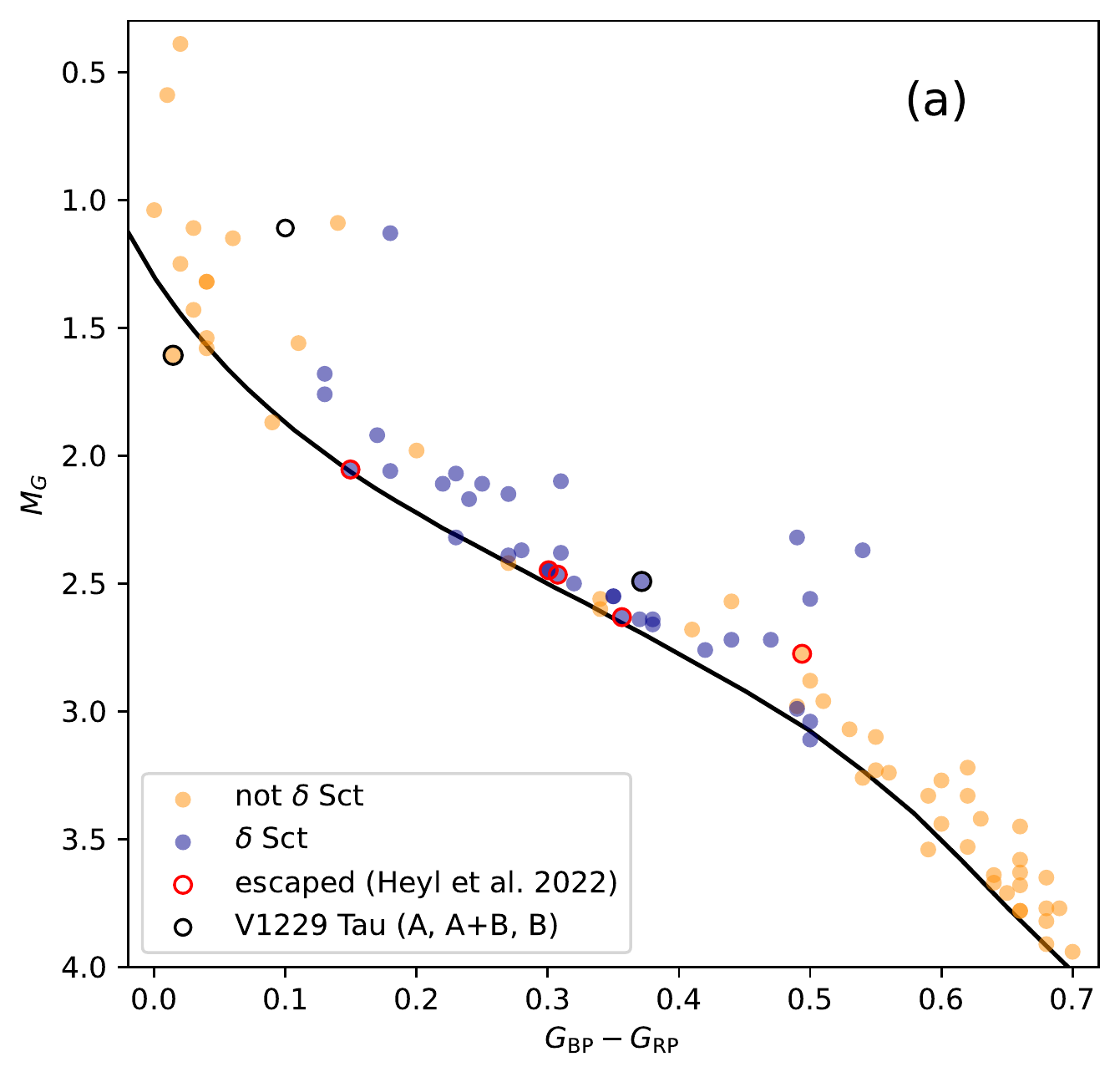}\\
\includegraphics[width=\linewidth]{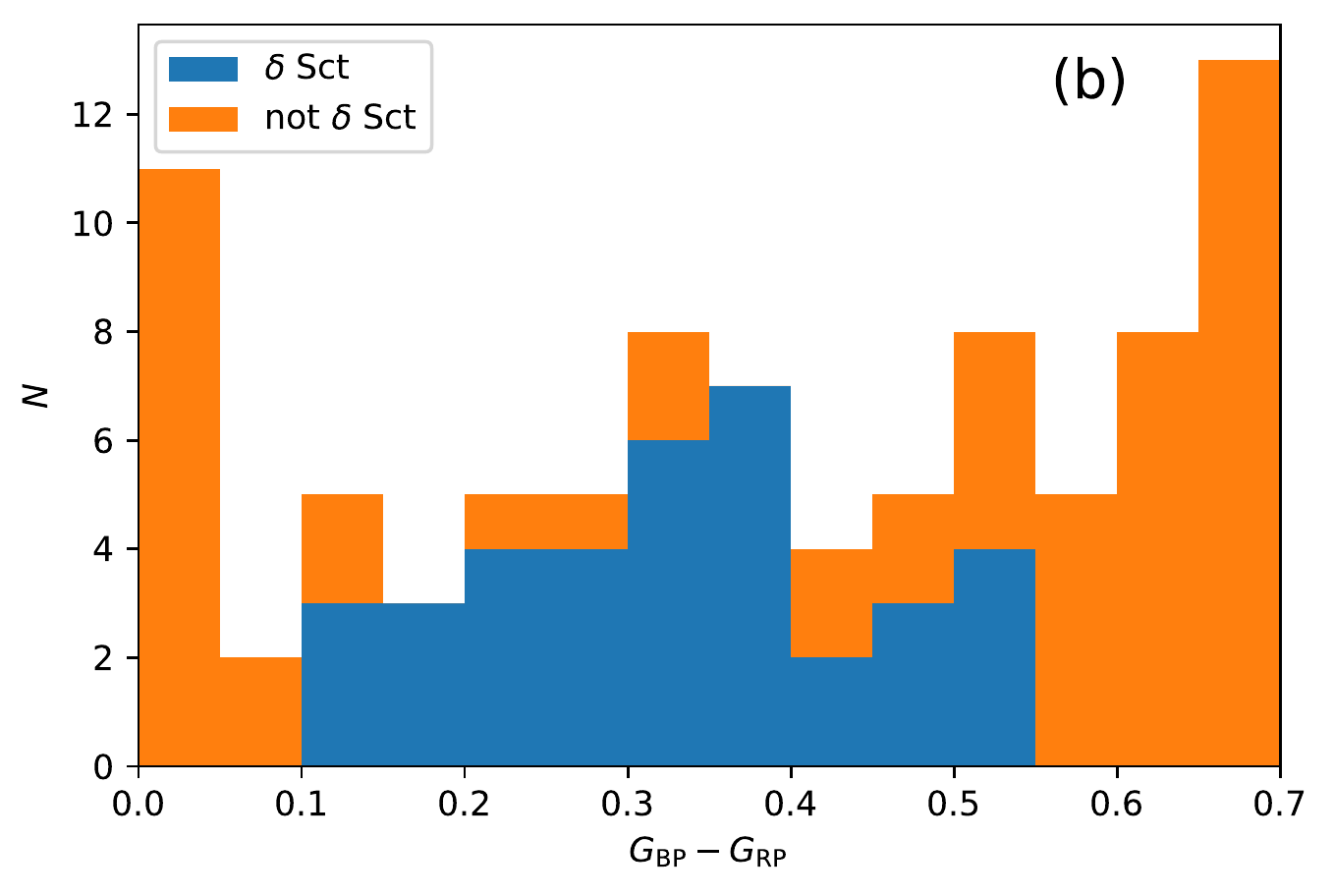}
\caption{Sample of 89 A and F stars in the Pleiades, of which 36 show \dsct\ pulsations (orange) and 53 do not (blue). (a)~Color-magnitude diagram, where red circular outlines are the five escaped members \citep{Heyl++2022}.
For the eclipsing binary V1229~Tau (HD~23642), black circular outlines show (from left to right) the A component (not pulsating), the combined system, and the B component (pulsating; see Section~\ref{sec:v1229-tau}).  
% black circular outlines are Am stars \citep{Renson+Manfroid2009}, and 
The black line is a PARSEC isoschrone (corrected for extinction and reddening; see Section~\ref{sec:sample}).  (b)~Histogram as a function of Gaia color.}
\label{fig:CMD-dsct}
\end{figure}

Figure~\ref{fig:CMD-dsct} shows the \dsct\ detections as a function of Gaia \bprp\ color index (without correcting for the reddening of the Pleiades, which is about 0.055; \citealt{Andrae++2018}).  We see in the CMD (Fig.~\ref{fig:CMD-dsct}a) and the accompanying histogram (Fig.~\ref{fig:CMD-dsct}b) that the pulsators lie within a strip that spans from about 0.10 to 0.55 in \bprp.  In this color range, the fraction of stars that pulsate is 36/50 ($72 \pm 6 \%$), and in the middle of the instability strip (0.20--0.40) it is 21/25 ($84 \pm 7 \%$).  This pulsator fraction is significantly higher than the 50--60\% found in the \kepler\ \dsct\ sample by \citet{Murphy2019}. 

% uncertainty is np.sqrt(frac*(1-frac)/N)

Although the Pleiades cluster is unusually rich in \dsct{}s (with peak amplitudes in the range 50--3000\,ppm, depending on the star), there are still several stars within the instability strip that are not pulsating (down to a sensitivity limit of 10--20\,ppm). One possible explanation is chemical peculiarity, which typically occurs in slow rotators because helium sinks out of the He\,{\sc ii} driving zone \citep{baglin++1973, deal++2020}. Slow rotation can be caused by tidal interactions with a binary companion \citep[e.g.,][and references therein]{fuller++2017}, which is thought to be responsible for the Am stars (``m'' for ``metallic-lined''; e.g., \citealt{abt1967,north++1998,debernardi++2000,Stateva++2012}). 

Eight stars in our sample were listed by \citet{Renson+Manfroid2009} as being Am stars (see column~13 in Table~\ref{tab:sample}).
One of these is the eclipsing binary V1229~Tau, for which \citet{Abt+Levato1978} gave the spectral type as A0\,Vp(Si) + Am,  indicating that the B component is an Am star.
Overall, seven of the Am stars in our sample have colors that place them within the \dsct\ instability strip, and five of these are pulsating.  The conclusion is that chemical peculiarity can only account for two of the non-pulsators in the Pleiades.  

\begin{figure}
\begin{center}
\includegraphics[width=\linewidth]{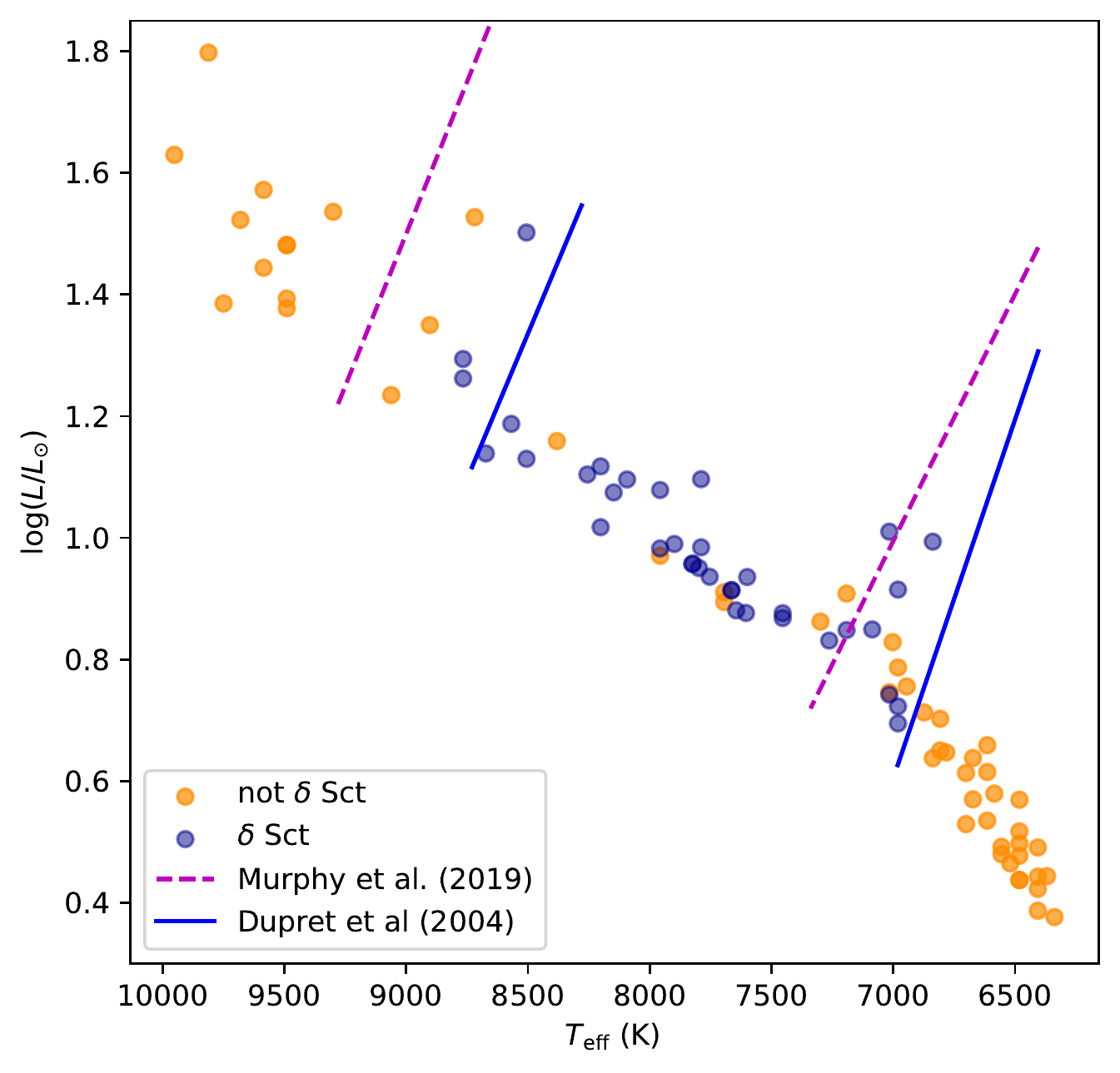}
\end{center}
\caption{H--R diagram of the Pleiades, corrected for extinction and reddening, showing the location of the \dsct\ pulsators.  Sloping lines indicate both the theoretical instability strip (solid blue lines; \citealt{Dupret2005}) and the observed strip from \kepler\ (dashed purple lines show the region in which at least 20\% of \kepler\ stars pulsate; \citealt{Murphy2019}).}
\label{fig:HRD}
\end{figure}

Figure~\ref{fig:HRD} shows our sample in an H--R diagram.  To construct this, we first corrected the observed Gaia photometry for extinction and reddening using the values given above.  We estimated effective temperatures from the de-reddened $\bprp$ colors using an updated version of Table~5 of \citet{Pecaut+Mamajek2013}\footnote{\url{http://www.pas.rochester.edu/~emamajek/EEM_dwarf_UBVIJHK_colors_Teff.txt}}.
We estimated approximate stellar luminosities from $G$ magnitudes and Gaia DR3 parallaxes, using $V$ bolometric corrections and $G-V$ colors from the same source.

The solid blue lines in Fig.~\ref{fig:HRD} show the theoretical instability strip calculated by \citet{Dupret2005}, \new{which used a typical solar composition ($X=0.7$, $Z=0.02$), a convective core overshooting parameter of $\alpha_{\rm ov} = 0.2$, and a solar-calibrated value for the mixing length parameter ($\alphaMLT=1.8$). 
With the caveat that \alphaMLT\ could have a different value for the Pleiades, we conclude that the observed \dsct\ strip for our sample is quite well matched to the theoretical calculations.}

%However, \citet{Murphy++2021-HD139614} showed that the choice for \alphaMLT\ was not important in the modelling of the young dSct star HD~139614, in the sense that the best-fitting oscillation frequencies did not depend on its value. \citet{Steindl++2021} showed that the instability regions are also insensitive to \alphaMLT, except near the end of the main sequence.}

The dashed purple lines in Fig.~\ref{fig:HRD} mark the instability strip for \dsct\ stars observed with \kepler\ \citep{Murphy2019}. The offset with respect to the Pleiades might come from a combination of: 
(1)~different $T_{\rm eff}$ scales being used; 
(2) having a homogeneous composition among Pleiades members, rather than the heterogeneous \kepler\ sample; 
(3) perhaps from having a slightly higher overall metallicity in the Pleiades; 
(4) from the Pleiades being young, as opposed to the \kepler\ sample where some stars that appeared near the ZAMS may be older stars of lower metallicity; and 
\new{(5)~the larger \kepler\ sample may give rise to more outliers at the hotter end of the distribution.}

\begin{figure}
\begin{center}
\includegraphics[width=\linewidth]{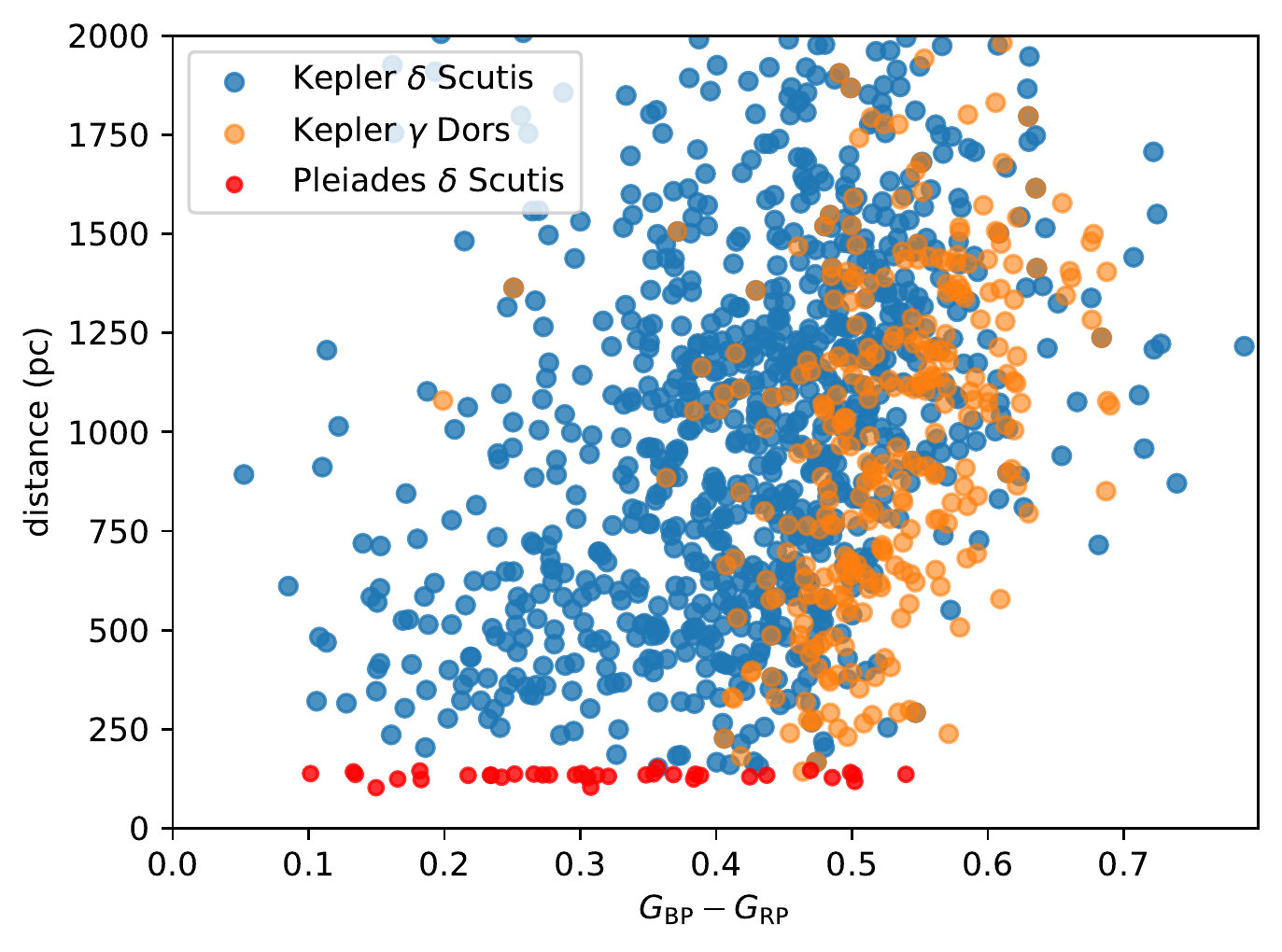}
\end{center}
\caption{The effect of reddening on the pulsation instability strip.  The \kepler\ samples of \dsct\ stars (from \citealt{Murphy2019}) and \gdor\ stars \citep{Li-Gang++2020} are restricted to Galactic latitude $b>10^\circ$. }
\label{fig:reddening}
\end{figure}

Figure~\ref{fig:reddening} shows the effect of reddening on the \dsct\ instability strip by plotting the distance to each star versus its Gaia color index.  The red points at the bottom show the Pleiades, and the blue points show \dsct\ stars detected by \kepler\ \citep{Murphy2019}.  As expected, the observed instability strip shifts to the red with increasing distance.  Note that we have restricted the \kepler\ sample to stars more than 10 degrees from the Galactic plane, in order to see the dependence on distance more clearly. \new{We see that the reddening in \bprp\ is approximately 0.15 magnitudes for every kpc in distance.}  We also show the sample of \gdor\ stars in the \kepler\ field studied by \citet{Li-Gang++2020}, again restricted to $b>10^\circ$ (orange points).  Overall, Fig.~\ref{fig:reddening} displays very nicely the effect of interstellar reddening on the pulsational instability strips. Note that we have not given a list of \gdor\ stars in the Pleiades, although many are certainly present, because having only a few \tess\ sectors often makes it difficult to distinguish unambiguously between gravity-mode pulsations and rotational modulation. 

%%%%%%%%%%%%%%%%%%%%%%%%%%%%%%%%%%%%%%%%%%%%%%%%%%%%%%%%%%%%%%%%
\section{Is there a \numax\ scaling relation for \texorpdfstring{$\delta$}{delta} Scuti stars?}
\label{sec:numax}

The quantity \numax\ is defined for solar-like oscillations as the centroid of the power envelope \citep{Kjeldsen+Bedding1995}.  
Solar-like oscillations are excited stochastically by near-surface convection, and the observed modes cover a broad range of frequencies centered at \numax.  It was suggested by \citet{Brown++1991} that when scaling from the Sun to other stars, \numax\ should be a fixed fraction of the acoustic cutoff frequency.  The latter is the frequency above which waves are no longer reflected at the surface, and is expected from simple arguments to scale as $g/\sqrt{\Teff}$. That line of reasoning underlies the scaling relation
\begin{equation}
    \numax \propto g/\sqrt{\Teff}, 
\end{equation}
which is widely used in the study of solar-like oscillations, although its physical basis is not well understood \citep{Belkacem++2011, Kjeldsen+Bedding2011, Chaplin+Miglio2013, Hekker2020, Zhou++2020}.

There have been suggestions that \numax\ is a useful observable for \dsct\ stars, given that it shows a correlation with \Teff\ \citep{Balona+Dziembowski2011,Barcelo-Forteza++2018, Bowman+Kurtz2018, Barcelo-Forteza++2020,Hasanzadeh++2021}.  However, the oscillation spectra of \dsct\ stars in the Pleiades (see Fig.~\ref{fig:amp-spectra}, which is ordered by color index) show the correlation to be weak or nonexistent for this sample of young main-sequence stars. 
In particular, we do not see a shift to higher radial orders with increasing \Teff, as predicted by theoretical models. Different theoretical treatments give slightly different predictions, but they all agree that we expect the excitation of higher radial order modes in \dsct\ stars as we move to higher temperatures within the instability strip \citep{Dziembowski1997, Houdek++1999, Pamyatnykh2000, Dupret2005, Houdek+Dupret2015, Xiong2016, Xiong2021}.  

\newtwo{It is also worth noting that there is no unambiguous definition for \numax\ in \dsct\ stars.  In a star with solar-like oscillations, the stochastic nature of the excitation and damping from convection leads to a power envelope of modes that is roughly Gaussian, with a well-defined maximum.  In many \dsct\ stars, on the other hand, the distribution of amplitudes is much less ordered (see Fig.~\ref{fig:amp-spectra}).  Therefore, even defining what is meant by \numax\ in \dsct\ stars is not straightforward.  One approach is to use the frequency of the strongest mode, $f_1$, which we have listed in Column~14 of Table~\ref{tab:sample}.  Given that many stars in Fig.~\ref{fig:amp-spectra} have several modes with similar amplitudes, this is clearly not an ideal metric. Not surprisingly, given the diversity in Fig.~\ref{fig:amp-spectra}, we found that plots of $f_1$ versus various stellar parameters (such as \Teff\ and \vsini) did not show any obvious correlations.}

The variety of oscillation spectra in Fig.~\ref{fig:amp-spectra} is quite remarkable, although there is also similarity between some stars.  
\new{\newtwo{To help guide the eye,} the vertical dotted lines in the figure show the approximate locations of the first 8 radial modes. \newtwo{These are based on the observed frequencies in the star V647~Tau, which has a particularly regular spectrum \citep[][Table~4]{Murphy++2022-K2-Pleiades}.}  

\newtwo{It is well-established that oscillation frequencies (and therefore also the large separation, \Dnu) scale as the square root of stellar density \citep[e.g.,][]{Aerts++2010-book}.}  Hence, we might expect \Dnu\ to vary substantially among the Pleiades \dsct\ stars, given their range of masses.  However, theoretical models of young main-sequence \dsct\ stars show that for fixed metallicity, \Dnu\ is remarkably constant across a wide range of masses (see Fig.4a of \citealt{Murphy++2021-HD139614} and Murphy et al., in preparation).  This makes the vertical lines in Fig.~\ref{fig:amp-spectra} quite useful for comparing the oscillation spectra.}

Some of the variations between stars could be attributed to differences in rotation, but it is difficult to see much in the way of systematic trends.  
\new{Furthermore, the distribution of \vsini\ values in the Pleiades, as noted in Sec.~\ref{sec:vsini}, is similar to that of A-type stars in general.}
On balance, our results seem to raise more questions than they answer.  On the one hand, the very high fraction of pulsators in the Pleiades means we are not left wondering why some pulsate and others do not.  On the other hand, we cannot explain why stars with similar properties have such different pulsation spectra, \new{although rotation presumably plays a role}.  An explanation for mode selection in \dsct\ stars remains as elusive as ever.

% \citet{Bowman+Kurtz2018} suggested that the \Teff--\numax scaling relation should be differentiated for different evolutionary stages.
 
% \citet{Michel2017} paper?

% In those stars where mode identification is possible, we see modes with $l=0$ and~1, but not $l=2$.  Again, this disagrees with expectations from theory that mode excitation is fairly weakly dependent on angular degree [ref]. Partial cancellation effects suggest that modes of $l=2$ should have amplitudes approximately half as strong as those of $l=0$ \citep{Aerts++2010-book}, but we do not observe this.

%%%%%%%%%%%%%%%%%%%%%%%%%%%%%%%%%%%%%%%%%%%%%%%%%%%%%%%%%%%%%%%%%%%%%%%%%%%%%
\section{Conclusions}

Using Gaia photometry and astrometry, we constructed a list of 89 probable members of the Pleiades with spectral types A and F.  We measured projected rotational velocities (\vsini) for 49 stars and confirmed that stellar rotation is a significant cause of the broadening of the main sequence in the color-magnitude diagram (Fig.~\ref{fig:CMD}b). Using time-series photometry from NASA's \tess\ Mission (plus one star observed by \kepler/K2), we detected \dsct\ pulsations in 36 stars.  Some stars suggested as being escaped members of the Pleiades by \citet{Heyl++2022} have similar pulsation properties to confirmed members, which supports their identification as former members.

The fraction of Pleiades stars in the middle of the instability strip that pulsate is unusually high (over 80\%), and their range of effective temperatures agrees well with theoretical models (Fig.~\ref{fig:HRD}).  On the other hand, the characteristics of the pulsation spectra are very varied and do not correlate very strongly with stellar temperature (Fig.~\ref{fig:amp-spectra}), calling into question the existence of a useful \numax\ relation for \dsct{}s, at least for young stars.  By including \dsct\ stars observed in the \kepler\ field (Fig.~\ref{fig:reddening}), we show that the instability strip is shifted to the red with increasing distance by interstellar reddening.  
In summary, this work demonstrates the power of combining observations with Gaia and \tess\ for studying pulsating stars in open clusters.

%%%%%%%%%%%%%%%%%%%%%%%%%%%%%%%%%%%%%%%%%%%%%%%%%%%%%%%%%%%%%%%%%%%%%%%%%%%%%
\acknowledgments

We thank the \tess\ team for making this research possible. The \tess\ data used in this paper can be found in MAST: \dataset[10.17909/t9-nmc8-f686]{http://dx.doi.org/10.17909/t9-nmc8-f686}.
We gratefully acknowledge support from the Australian Research Council through Discovery Project DP210103119, Future Fellowship FT210100485 and Laureate Fellowship FL220100117, and from the Danish National Research Foundation (Grant DNRF106) through its funding for the Stellar Astrophysics Centre (SAC). 
D.H. acknowledges support from the Alfred P. Sloan Foundation and the National Aeronautics and Space Administration (80NSSC21K0784).
This research made use of {\sc Lightkurve}, a Python package for \kepler\ and \tess\ data analysis \citep{lightkurve2018}.
This work made use of several publicly available {\tt python} packages: {\tt astropy} \citep{astropy:2013,astropy:2018}, 
{\tt lightkurve} \citep{lightkurve2018},
{\tt matplotlib} \citep{matplotlib2007}, 
{\tt numpy} \citep{numpy2020}, and 
{\tt scipy} \citep{scipy2020}.

This work has made use of data from the European Space Agency (ESA) mission Gaia (\url{https://www.cosmos.esa.int/gaia}), processed by the Gaia Data Processing and Analysis Consortium (DPAC, \url{https://www.cosmos.esa.int/web/gaia/dpac/consortium}). Funding for the DPAC has been provided by national institutions, in particular the institutions participating in the Gaia Multilateral Agreement. 
\new{We thank Konstanze Zwintz, Dennis Stello and the referee for useful comments on the paper.}

\facility{TESS}

\clearpage
% table is truncated if placed last! :-(
\startlongtable
\begin{longrotatetable}
\begin{deluxetable*}{rrrrDDDccccccDD}
\tablenum{1}
\tablecaption{Sample of A and F stars in the Pleiades.
\newtwo{Column~8: 1 indicates an escaped member \citep{Heyl++2022}.}
\new{Column~9: 1 indicates a spectroscopic binary \citep[\newtwo{SB;}][]{Torres++2021}.}
Column~11: source for \vsini\ is 1 (this work) or 2 (Gaia DR3).
Column~12: 1 indicates \dsct\ star.
Column~13: 1 indicates Am star \citep{Renson+Manfroid2009}.
\new{Columns~14 and~15: frequency and amplitude of the strongest \dsct\ mode.}
For V1229~Tau A \& B, the magnitudes and colors (columns 5--7) are estimates (see Sec.~\ref{sec:v1229-tau}).
\label{tab:sample}}
\tablewidth{0pt}
\tablehead{
\colhead{Name} & \colhead{HD} & \colhead{\newtwo{HIP}} & \colhead{TIC} &
\twocolhead{$G$} & \twocolhead{$M_G$} & \twocolhead{\bprp} & \colhead{\newtwo{Esc.}} & \colhead{SB} & 
\twocolhead{$\vsini$} & \colhead{$\delta$\,Sct} & \colhead{Am} & \twocolhead{$f_1$} & \twocolhead{$a_1$} \\
\colhead{} & \colhead{} & \colhead{} & \colhead{} &
\twocolhead{} & \twocolhead{} & \twocolhead{} & \colhead{} & \colhead{} &
\colhead{\kms} & \colhead{src} & \colhead{} & \colhead{} & \twocolhead{\cd} & \twocolhead{ppt}
}
\decimalcolnumbers
\startdata
 & 22578 & 17000 & 113956708 & 6.70 & 1.04 & 0.00 & 0 & 0 & 242 $\pm$ 22 & 1 & 0 & 0 & . & . \\
24 Tau & 23629 &  & 405484171 & 6.30 & 0.59 & 0.01 & 0 & 0 & 184 $\pm$ 15 & 1 & 0 & 0 & . & . \\
V1229 Tau A & 23642 & 17704 & 125754991 & 7.32 & 1.61 & 0.01 & 0 & 1 &  &  & 0 & 0 & . & . \\
 & 23410 & 17572 & 67830155 & 6.90 & 1.25 & 0.02 & 0 & 0 & 180 $\pm$ 9 & 1 & 0 & 0 & . & . \\
 & 23950 & 17921 & 440695282 & 6.04 & 0.39 & 0.02 & 0 & 0 & 133 $\pm$ 7 & 1 & 0 & 0 & . & . \\
 & 24899 & 18559 & 149980785 & 7.20 & 1.43 & 0.03 & 0 & 1 & 67 $\pm$ 2 & 1 & 0 & 0 & . & . \\
 & 23568 & 17664 & 405484416 & 6.80 & 1.11 & 0.03 & 0 & 0 & 188 $\pm$ 5 & 2 & 0 & 0 & . & . \\
 & 22614 & 17034 & 427545204 & 7.10 & 1.54 & 0.04 & 0 & 0 & 114 $\pm$ 2 & 1 & 0 & 0 & . & . \\
 & 23631 &  & 440681316 & 7.29 & 1.58 & 0.04 & 0 & 1 & 13 $\pm$ 5 & 1 & 0 & 1 & . & . \\
 & 23632 & 17692 & 440681358 & 7.00 & 1.32 & 0.04 & 0 & 0 & 194 $\pm$ 3 & 2 & 0 & 0 & . & . \\
 & 23913 & 17892 & 440691760 & 7.00 & 1.32 & 0.04 & 0 & 0 & 205 $\pm$ 21 & 1 & 0 & 0 & . & . \\
 & 23964 & 17923 & 440695975 & 6.81 & 1.15 & 0.06 & 0 & 1 & 5 $\pm$ 1 & 1 & 0 & 0 & . & . \\
 & 23948 &  & 35159593 & 7.55 & 1.87 & 0.09 & 0 & 0 & 115 $\pm$ 1 & 2 & 0 & 0 & . & . \\
 & 22637 & 17043 & 113981021 & 7.27 & 1.56 & 0.11 & 0 & 1 & 131 $\pm$ 10 & 1 & 0 & 0 & . & . \\
 & 23872 &  & 346626001 & 7.53 & 1.76 & 0.13 & 0 & 0 & 246 $\pm$ 5 & 2 & 1 & 0 & 62.01 & 0.09 \\
 & 23489 &  & 125736946 & 7.35 & 1.68 & 0.13 & 0 & 0 & 124 $\pm$ 2 & 1 & 1 & 0 & 22.64 & 0.08 \\
 & 24076 & 17999 & 35205647 & 6.93 & 1.09 & 0.14 & 0 & 0 &  &  & 0 & 0 & . & . \\
 & 21062 & 15902 & 29058513 & 7.11 & 2.05 & 0.15 & 1 & 0 & 118 $\pm$ 1 & 2 & 1 & 0 & 47.64 & 1.05 \\
 & 23336 & 17547 & 385554826 & 7.40 & 1.92 & 0.17 & 0 & 0 & 243 $\pm$ 3 & 2 & 1 & 0 & 31.33 & 0.67 \\
 & 23763 & 17791 & 35156298 & 6.94 & 1.13 & 0.18 & 0 & 0 & 116 $\pm$ 2 & 1 & 1 & 0 & 33.28 & 0.10 \\
 & 23155 & 17403 & 405483425 & 7.51 & 2.06 & 0.18 & 0 & 0 & 205 $\pm$ 5 & 1 & 1 & 0 & 32.92 & 1.05 \\
 & 24178 &  & 84336172 & 7.65 & 1.98 & 0.20 & 0 & 0 & 214 $\pm$ 13 & 1 & 0 & 0 & . & . \\
V650 Tau & 23643 &  & 440681425 & 7.75 & 2.11 & 0.22 & 0 & 0 & 238 $\pm$ 7 & 1 & 1 & 0 & 32.64 & 2.97 \\
 & 23886 &  & 346626099 & 7.96 & 2.32 & 0.23 & 0 & 0 & 125 $\pm$ 2 & 2 & 1 & 0 & 56.62 & 1.94 \\
 & 23852 &  & 440691730 & 7.71 & 2.07 & 0.23 & 0 & 0 & 148 $\pm$ 4 & 1 & 1 & 0 & 17.77 & 1.59 \\
 & 23388 & 17552 & 67828699 & 7.73 & 2.17 & 0.24 & 0 & 0 & 204 $\pm$ 7 & 1 & 1 & 0 & 14.14 & 0.51 \\
 & 23402 &  & 67830321 & 7.80 & 2.11 & 0.25 & 0 & 0 & 248 $\pm$ 6 & 1 & 1 & 0 & 40.86 & 0.42 \\
V1187 Tau & 23194 &  & 405483707 & 8.04 & 2.39 & 0.27 & 0 & 0 & 37 $\pm$ 1 & 1 & 1 & 1 & 53.18 & 1.46 \\
 & 23924 &  & 440695768 & 8.09 & 2.42 & 0.27 & 0 & 0 &  &  & 0 & 1 & . & . \\
 & 23409 &  & 385589694 & 7.83 & 2.15 & 0.27 & 0 & 0 & 213 $\pm$ 5 & 1 & 1 & 0 & 32.31 & 1.95 \\
 & 23430 & 17583 & 385558439 & 8.02 & 2.37 & 0.28 & 0 & 0 & 130 $\pm$ 3 & 1 & 1 & 0 & 44.99 & 0.86 \\
 & 23863 &  & 346626294 & 8.10 & 2.45 & 0.30 & 0 & 0 & 163 $\pm$ 6 & 1 & 1 & 0 & 41.47 & 1.28 \\
 & 17962 & 13522 & 77568727 & 8.15 & 2.45 & 0.30 & 1 & 0 & 152 $\pm$ 2 & 2 & 1 & 0 & 39.35 & 0.57 \\
 & 20655 & 15552 & 402366726 & 7.55 & 2.47 & 0.31 & 1 & 0 & 146 $\pm$ 2 & 2 & 1 & 0 & 34.88 & 1.38 \\
 & 23361 &  & 385552144 & 8.02 & 2.38 & 0.31 & 0 & 0 & 219 $\pm$ 8 & 1 & 1 & 0 & 32.66 & 0.97 \\
V1228 Tau & 23628 &  & 125754823 & 7.63 & 2.10 & 0.31 & 0 & 0 &  &  & 1 & 0 & 32.54 & 0.74 \\
 & 21744 & 16407 & 46476992 & 8.09 & 2.50 & 0.32 & 0 & 0 & 130 $\pm$ 3 & 1 & 1 & 0 & 41.78 & 0.81 \\
 & 23664 & 17729 & 125754460 & 8.27 & 2.56 & 0.34 & 0 & 0 & 96 $\pm$ 2 & 1 & 0 & 0 & . & . \\
 & 23610 & 17694 & 440681752 & 8.12 & 2.60 & 0.34 & 0 & 1 & 26 $\pm$ 1 & 1 & 0 & 1 & . & . \\
V624 Tau & 23156 &  & 405483817 & 8.20 & 2.55 & 0.35 & 0 & 0 &  &  & 1 & 0 & 39.03 & 1.67 \\
V647 Tau & 23607 &  & 405484188 & 8.24 & 2.55 & 0.35 & 0 & 0 & 19 $\pm$ 1 & 1 & 1 & 1 & 38.38 & 1.47 \\
 & 23323 &  & 385553714 & 8.55 & 2.63 & 0.36 & 1 & 0 & 123 $\pm$ 1 & 2 & 1 & 0 & 20.81 & 2.49 \\
 & 24711 & 18431 & 14111056 & 8.30 & 2.64 & 0.37 & 0 & 0 & 138 $\pm$ 3 & 1 & 1 & 0 & 42.82 & 0.79 \\
V1229 Tau B & 23642 & 17704 & 125754991 & 8.20 & 2.49 & 0.37 & 0 & 1 &  &  & 1 & 1 & 21.89 & 0.15 \\
 & 23246 &  & 348639016 & 8.12 & 2.64 & 0.38 & 0 & 0 &  &  & 1 & 0 & 23.31 & 0.19 \\
 & 23791 &  & 440690782 & 8.34 & 2.66 & 0.38 & 0 & 0 &  &  & 1 & 1 & 20.89 & 0.38 \\
V1210 Tau & 23585 &  & 405484093 & 8.33 & 2.68 & 0.41 & 0 & 0 & 108 $\pm$ 3 & 1 & 0 & 0 & . & . \\
 & 21510 & 16217 & 405461432 & 8.33 & 2.76 & 0.42 & 0 & 0 &  &  & 1 & 0 & 21.75 & 0.45 \\
 & 23479 &  & 385589599 & 8.23 & 2.57 & 0.44 & 0 & 0 &  &  & 0 & 0 & . & . \\
 & 23028 & 17325 & 114083179 & 8.36 & 2.72 & 0.44 & 0 & 0 & 68 $\pm$ 1 & 1 & 1 & 0 & . & . \\
 & 23325 &  & 385509282 & 8.55 & 2.72 & 0.47 & 0 & 0 & 85 $\pm$ 1 & 2 & 1 & 1 & 23.22 & 0.64 \\
 & 23157 & 17401 & 67768222 & 7.86 & 2.32 & 0.49 & 0 & 0 & 58 $\pm$ 1 & 1 & 1 & 0 & 32.46 & 0.55 \\
V1225 Tau & 22702 &  & 427580304 & 8.75 & 2.98 & 0.49 & 0 & 0 & 137 $\pm$ 6 & 1 & 0 & 0 & . & . \\
 & 23488 & 17625 & 125736216 & 8.65 & 2.99 & 0.49 & 0 & 1 & 18 $\pm$ 1 & 1 & 1 & 0 & 22.04 & 0.88 \\
 & 34027 & 24808 & 82969878 & 8.85 & 2.77 & 0.49 & 1 & 0 &  &  & 0 & 0 & . & . \\
 & 23375 &  & 385552372 & 8.55 & 2.88 & 0.50 & 0 & 0 &  &  & 0 & 0 & . & . \\
V534 Tau & 23567 &  & 405484574 & 8.48 & 3.11 & 0.50 & 0 & 0 &  &  & 1 & 0 & 38.71 & 1.06 \\
 & 23733 &  & 35155873 & 8.21 & 2.56 & 0.50 & 0 & 1 & 166 $\pm$ 25 & 1 & 1 & 0 & 18.26 & 0.16 \\
 & 22146 &  & 26126738 & 8.79 & 3.04 & 0.50 & 0 & 0 &  &  & 1 & 0 & 11.13 & 0.05 \\
 & 23290 & 17481 & 67788829 & 8.63 & 2.96 & 0.51 & 0 & 0 &  &  & 0 & 0 & . & . \\
 & 24132 & 18050 & 84331341 & 8.77 & 3.07 & 0.53 & 0 & 0 &  &  & 0 & 0 & . & . \\
 & 23326 &  & 67829720 & 8.89 & 3.26 & 0.54 & 0 & 0 & 19 $\pm$ 1 & 1 & 0 & 0 & . & . \\
 & 23512 &  & 61139371 & 8.04 & 2.37 & 0.54 & 0 & 0 & 170 $\pm$ 8 & 1 & 1 & 0 & 53.64 & 0.05 \\
 & 23792 &  & 440690206 & 8.31 & 3.10 & 0.55 & 0 & 1 & 164 $\pm$ 11 & 1 & 0 & 0 & . & . \\
 & 23289 & 17497 & 67787772 & 8.89 & 3.23 & 0.55 & 0 & 0 & 26 $\pm$ 1 & 1 & 0 & 0 & . & . \\
 &  & 16423 & 26078071 & 8.78 & 3.24 & 0.56 & 0 & 0 &  &  & 0 & 0 & . & . \\
 & 24655 &  & 14109779 & 8.98 & 3.54 & 0.59 & 0 & 0 & 22 $\pm$ 1 & 1 & 0 & 0 & . & . \\
 & 23912 &  & 440691379 & 9.03 & 3.33 & 0.59 & 0 & 0 & 151 $\pm$ 4 & 1 & 0 & 0 & . & . \\
 & 22887 & 17225 & 114060256 & 9.07 & 3.44 & 0.60 & 0 & 0 &  &  & 0 & 0 & . & . \\
 & 23133 &  & 114166637 & 8.89 & 3.27 & 0.60 & 0 & 0 & 122 $\pm$ 22 & 1 & 0 & 0 & . & . \\
 & 23351 &  & 385552643 & 8.90 & 3.22 & 0.62 & 0 & 1 &  &  & 0 & 0 & . & . \\
 & 23511 &  & 125736995 & 9.20 & 3.53 & 0.62 & 0 & 0 & 30 $\pm$ 1 & 1 & 0 & 0 & . & . \\
 & 24086 &  & 84331854 & 9.01 & 3.33 & 0.62 & 0 & 0 &  &  & 0 & 0 & . & . \\
 & 22977 & 17289 & 114084434 & 9.06 & 3.42 & 0.63 & 0 & 0 &  &  & 0 & 0 & . & . \\
 & 24302 & 18154 & 427735820 & 9.34 & 3.67 & 0.64 & 0 & 0 &  &  & 0 & 0 & . & . \\
 & 23513 &  & 61145701 & 9.30 & 3.64 & 0.64 & 0 & 0 & 32 $\pm$ 1 & 1 & 0 & 0 & . & . \\
 & 23584 &  & 405484278 & 9.38 & 3.71 & 0.65 & 0 & 0 & 82 $\pm$ 2 & 1 & 0 & 0 & . & . \\
 & 23312 & 17511 & 67788288 & 9.36 & 3.63 & 0.66 & 0 & 0 &  &  & 0 & 0 & . & . \\
 &  & 17125 & 353928999 & 9.50 & 3.78 & 0.66 & 0 & 0 & 85 $\pm$ 2 & 1 & 0 & 0 & . & . \\
 & 23514 &  & 61145611 & 9.31 & 3.58 & 0.66 & 0 & 0 &  &  & 0 & 0 & . & . \\
 &  & 18544 & 14177821 & 9.29 & 3.78 & 0.66 & 0 & 0 & 72 $\pm$ 2 & 1 & 0 & 0 & . & . \\
 & 23732 &  & 35155396 & 9.12 & 3.45 & 0.66 & 0 & 0 & 23 $\pm$ 1 & 1 & 0 & 0 & . & . \\
 & 23061 &  & 258067594 & 9.37 & 3.68 & 0.66 & 0 & 0 &  &  & 0 & 0 & . & . \\
SAO 93581 &  &  & 67789284 & 9.30 & 3.65 & 0.68 & 0 & 0 &  &  & 0 & 0 & . & . \\
 & 23975 &  & 35204900 & 9.52 & 3.82 & 0.68 & 0 & 0 &  &  & 0 & 0 & . & . \\
 &  & 16639 & 46538779 & 9.43 & 3.77 & 0.68 & 0 & 0 &  &  & 0 & 0 & . & . \\
 & 23352 &  & 385552619 & 9.57 & 3.91 & 0.68 & 0 & 0 & 34 $\pm$ 1 & 1 & 0 & 0 & . & . \\
 & 23158 &  & 67768242 & 9.43 & 3.77 & 0.69 & 0 & 1 & 40 $\pm$ 1 & 1 & 0 & 0 & . & . \\
 & 24463 &  & 348769726 & 9.60 & 3.94 & 0.70 & 0 & 0 &  &  & 0 & 0 & . & . \\
 
\enddata
\end{deluxetable*}
\end{longrotatetable}

\clearpage

\ifarxiv
    
 % for arxiv version
\else
    \bibliography{references,Am_refs,AWMref}{}
    \bibliographystyle{aasjournal}
\fi

\end{document}